\documentclass[fleqn,10pt]{wlscirep}
\usepackage[utf8]{inputenc}
\usepackage[T1]{fontenc}
\usepackage{bm}
\usepackage{siunitx}
\title{Topology Optimization of Random Memristors for Input-Aware Dynamic SNN}

\author[1,2]{Bo Wang}
\author[1,2]{Shaocong Wang}
\author[1,2]{Ning Lin}
\author[1,3,4]{Yi Li}
\author[1,2]{Yifei Yu}
\author[1,2]{Yue Zhang}
\author[1,2]{Jichang Yang}
\author[1,2]{Xiaoshan Wu}
\author[1,2]{Yangu He}
\author[1,2]{Songqi Wang}
\author[3,6]{Rui Chen}
\author[5,6]{Guoqi Li}
\author[1]{Xiaojuan Qi}
\author[1,2*]{Zhongrui Wang}
\author[3,4,6*]{Dashan Shang}

\affil[1]{Department of Electrical and Electronic Engineering, the University of Hong Kong, Hong Kong, China}
\affil[2]{ACCESS – AI Chip Center for Emerging Smart Systems, InnoHK Centers, Hong Kong Science Park, Hong Kong, China}
\affil[3]{Key Lab of Fabrication Technologies for Integrated Circuits, Institute of Microelectronics, Chinese Academy of Sciences, Beijing 100029, China}
\affil[4]{Laboratory of Microelectronic Devices \& Integrated Technology, Institute of Microelectronics, Chinese Academy of Sciences, Beijing 100029, China}
\affil[5]{Key Laboratory of Brain Cognition and Brain-inspired Intelligence Technology, Institute of Automation, Chinese Academy of Sciences, Beijing 100190, China}
\affil[6]{University of Chinese Academy of Sciences, Beijing 100049, China}

\affil[*]{e-mail: zrwang@eee.hku.hk; shangdaashan@ime.cas.ac}

\begin{abstract}

There is unprecedented development in machine learning, exemplified by recent large language models (GPT4) and world simulators (SORA), which are artificial neural networks (ANNs) running on digital computers.
However, they still cannot parallel human brains in terms of energy efficiency and the streamlined adaptability to inputs of different difficulties, due to differences in signal representation, optimization, run-time reconfigurability, and hardware architecture.
To address these fundamental challenges, we introduce \textbf{pr}uning optimization for \textbf{i}nput-aware dynamic \textbf{me}mristive spiking neural network (PRIME). 
Signal representation-wise, PRIME employs leaky integrate-and-fire neurons to emulate the brain's inherent spiking mechanism. Drawing inspiration from the brain's structural plasticity, PRIME optimizes the topology of a random memristive SNN, without expensive memristor conductance fine-tuning. For runtime reconfigurability, inspired by the brain's dynamic adjustment of computational depth, PRIME employs an input-aware dynamic early stop policy to minimize latency during inference, thereby boosting energy efficiency without compromising performance. Architecture-wise, PRIME leverages memristive in-memory computing, mirroring the brain and mitigating the von Neumann bottleneck.
We validated our system using a 40 nm 256 Kb memristor-based in-memory computing macro on neuromorphic image classification and image inpainting. Our results demonstrate the classification accuracy and Inception Score (IS) are comparable to the software baseline, while achieving 37.83\(\times\) and 62.50\(\times\) improvements in energy efficiency. Furthermore, we attained 77.0\% and 12.5\% computational load savings with minimal reduction in performance. The system also exhibits robustness against stochastic synaptic noise of analogue memristors. 
Our software-hardware co-designed model paves the way to future brain-inspired neuromorphic computing of brain-like energy efficiency and adaptivity.

\end{abstract}
\begin{document}

\flushbottom
\maketitle

\thispagestyle{empty}

\begin{figure}[!t]
\centering
\includegraphics[width=0.9\linewidth]{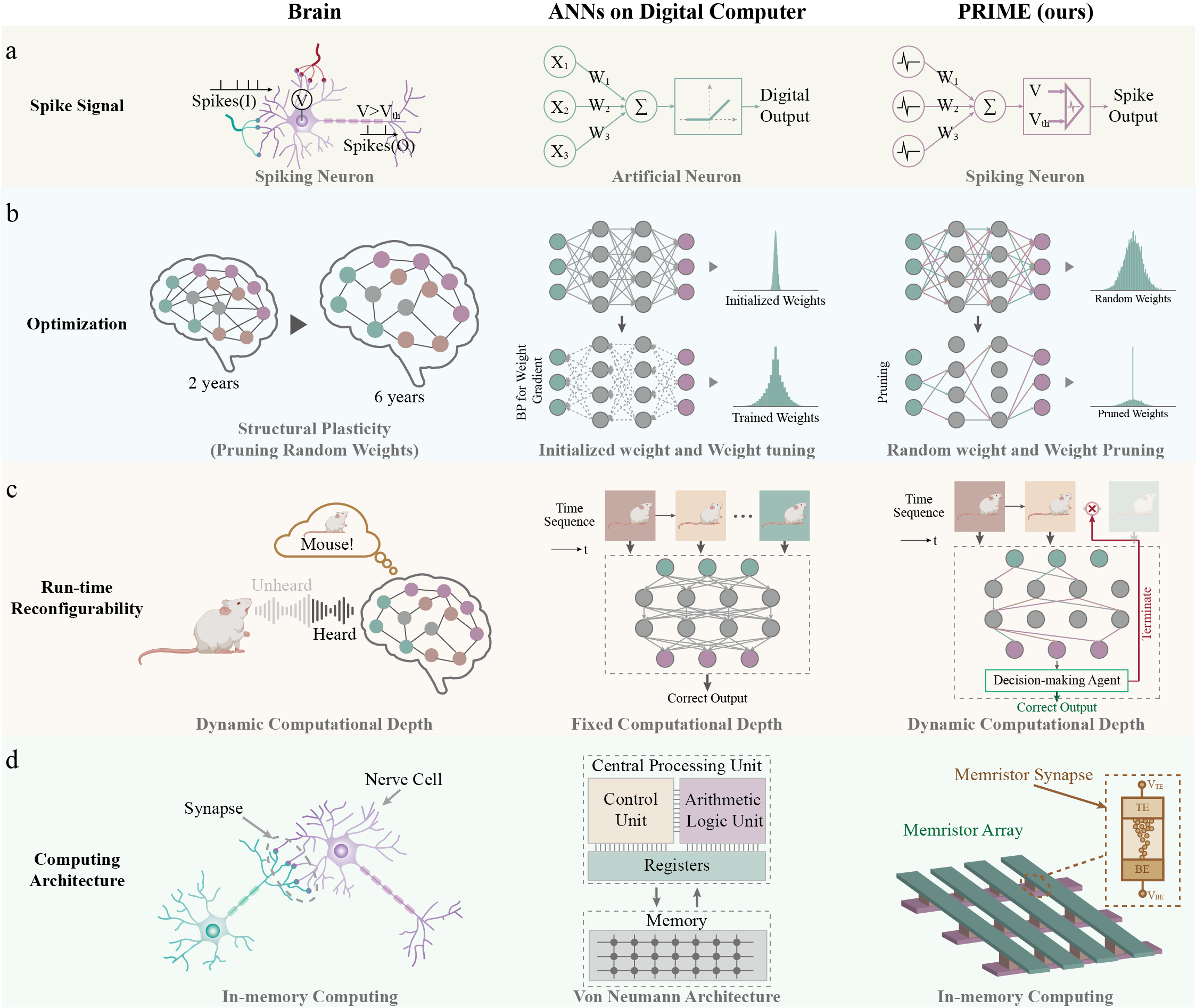}
\caption{\textbf{Brain-inspired topology optimization for input-aware dynamic SNN on memristors.}
\textbf{a,} Comparison of the information representation in human brain, the artificial neuron model of conventional static ANNs, and the spiking neuron model of PRIME. Both the human brain and PRIME encode information with spikes, whereas ANNs do not.
\textbf{b,} Comparison of the optimization scheme in human brain with structural plasticity, weight-trained neural network, and topology-optimized neural network. The human brain and PRIME optimize network topology instead of relying on fine-tuning of synaptic weights, as seen in conventional ANNs.
\textbf{c,} Comparison of run-time reconfigurability in human brain with dynamic computational depth, conventional static ANNs, and PRIME. The human brain and PRIME feature dynamic computational depth and dynamically adapt to new stimuli for reducing computational costs. In contrast, conventional ANNs are of fixed computational depth that is constant to inputs of different difficulties.
\textbf{d,} Comparison of the hardware architecture in human brain, the digital hardware implementing conventional ANNs, and memristive neuromorphic system on PRIME. The human brain and PRIME utilize in-memory computing, which collocates processing and memory in biological synapses and memristors, respectively, thereby enhancing energy efficiency. In contrast, digital computers based on the Von Neumann architecture, separate storage and computing.}
\label{fig1}
\end{figure}

\section*{Introduction}


Machine learning with artificial neural networks (ANNs) has undergone significant advancements in recent years\cite{lecun2015deep, philips_1, yuchao_1}, as demonstrated by the development of sophisticated large language models\cite{chang2023survey, zhao2023survey} such as GPT4 and advanced world simulators\cite{videoworldsimulators2024} like SORA. These models, operating on digital computers, exhibit human-like capabilities and are steps toward the long-term goal of artificial general intelligence (AGI).


Despite these achievements, a discernible discrepancy in performance persists compared to the human brain, especially in terms of energy efficiency\cite{yuchao_2, JJYang_1, philips_2, huaqiang_1, luwei_1} and adapting to inputs of different difficulties by dynamically allocating computing resources\cite{dynamic_com1, dynamic_com2, dynamic_com3, dynamic_com4, dynamic_com5}. This gap can be attributed to fundamental differences in signal representation, optimization, run-time reconfigurability, and architecture.

In terms of  signal representation, the human brain utilizes spikes\cite{llinas1982electrophysiology, hahn2019portraits} for information representation (Fig. ~\ref{fig1}a, left). These spikes are sparse, robust to signal noise, and enable the brain to perform advanced cognitive tasks at a minimal energy of only 20 watts\cite{wozniak2020deep, mehonic2022brain}. In contrast, mainstream AI systems employ digital-valued computation instead of the spiking mechanisms (Fig. ~\ref{fig1}a, middle). 

In terms of optimization, a major step in brain's development is initially characterized by an abundance of random synaptic connections. This configuration undergoes structural plasticity\cite{paolicelli2011synaptic, faust2021mechanisms, sellgren2019increased}, a process that entails the elimination of less important connections while retaining informative ones (Fig. ~\ref{fig1}b, left). In contrast, current AI predominantly concentrate on the meticulous tuning of synaptic weights (Fig. ~\ref{fig1}b, middle) that does not work well on emerging synaptic devices. 

In terms of run-time reconfigurability, the brain dynamically reallocates its computational resources by adjusting the depth of computation in sequential decision-making tasks according to task demands\cite{dynamic_com1, dynamic_com2, dynamic_com3, dynamic_com4, dynamic_com5}. When solving complex tasks, the human brain estimates future consequences by integrating past experiences\cite{dynamic_com2, dynamic_com3, dynamic_com4}, including accumulated scores\cite{dynamic_com3}, the sequential interdependence of previous inputs\cite{dynamic_com4}, and etc\cite{dynamic_com2}, to determine the appropriate termination point for making a final decision. This dynamic computational depth enables the brain to balance efficiency and accuracy within the constraints of limited computational budget\cite{dynamic_com1, dynamic_com5}. Nevertheless, major conventional ANNs remain static, which works on entire inputs before decision making (Fig. ~\ref{fig1}c, middle).

In terms of hardware architecture,  nerve cells communicate with each other through synaptic connections (Fig. ~\ref{fig1}d, left). The latter both store information as synaptic strength and perform computation (i.e. modulating signal transmission) right at where the information is stored\cite{knoblauch2010memory, abbott2004synaptic}. This mechanism is highly energy efficient as no data movement is needed, while being parallel. Conversely, mainstream AI hardware relies on the digital von Neumann architecture\cite{wang2020resistive, huaqiang_2, luwei_2}, which physically separates memory from computing, incurring large energy and time consumption due to massive data shuttling.

To address these differences, we proposed \textbf{pr}uning optimization for \textbf{i}nput-aware dynamic \textbf{me}mristive spiking neural network (PRIME). 

In terms of signal representation, PRIME is a spiking neural network (SNN) with brain-inspired leaky integrate-and-fire neurons to process and propagate information (Fig. ~\ref{fig1}a, right). This spiking mechanism enhances energy efficiency with emerging neuromorphic hardware.

In terms of optimization, PRIME implements structural plasticity in a manner akin to human brain. According to strong lottery ticket hypothesis theory\cite{ramanujan2020s, LTH_2}, a sub-net sampled from a supernet with sufficient random weights can achieve competitive accuracy relative to the target network with well-optimized weights. Here PRIME leverages the inherent programming stochasticity of memristive synapses for initiating random weights, followed by topology optimization (Fig. ~\ref{fig1}b, right) to preserve informative synapses while eliminating those considered less critical. This topology optimization approach naturally avoids the energy-intensive and time-consuming memristor conductance tuning.

In terms of run-time reconfigurability, inspired by the the dynamic computational depth of human brain, PRIME implements the input-aware dynamic latency for SNN during the inference phase. Since the latency (represented by the number of timesteps) in SNN significantly influences performance and directly correlates with energy consumption in hardware\cite{li2023input, li2024seenn, li2023unleashing}, we introduce an agent to compute time-wise SNN output confidence, which dynamically adjusts timesteps for each input sample (Fig. ~\ref{fig1}c, right). This strategy adapts to inputs of different difficulties at minimal accuracy loss and computational load.

In terms of hardware architecture, PRIME employs neuromorphic in-memory computing paradigm analogous to the brain, utilizing memristor synapses to store synaptic weights and modulate signal transmission between spiking neurons (Fig. ~\ref{fig1}d, right). The integration of memory and computing in neuromorphic hardware substantially enhances parallelism and reduces energy consumption compared with the traditional von Neumann architecture (Fig. ~\ref{fig1}d, middle). Additionally, we exploit the inherent electroforming stochasticity of memristors to produce large-scale and cost-effective true random weights, turning it into an advantage through our topology optimization strategy.

Our research validates this system on two tasks: event image classification and image inpainting, executed on a hybrid analogue-digital system with a 40nm 256K memristor-based in-memory computing core. In the N-MNIST classification task, PRIME achieves comparable accuracy to software baseline but with 37.83\(\times\) improvements in energy efficiency and 67.6\% reduction in calculation cost. For image inpainting, our system parallels the reconstruction loss and Inception Score (IS)\cite{salimans2016improved} of software baseline, showcasing its ability to generate diverse and high-quality images. Additionally, it achieves approximately 62.50\(\times\) reductions in energy consumption compared to software baseline on conventional digital hardware and 12.5\% computational load savings with minimal performance degradation. Furthermore, PRIME is robust to synaptic noise. PRIME paves the way for future brain-inspired, energy-efficient, and low-latency neuromorphic computing paradigm.

\section*{PRIME: topology pruning optimization for input-aware dynamic memristive SNN}

\begin{figure}[!t]
\centering
\includegraphics[width=0.9\linewidth]{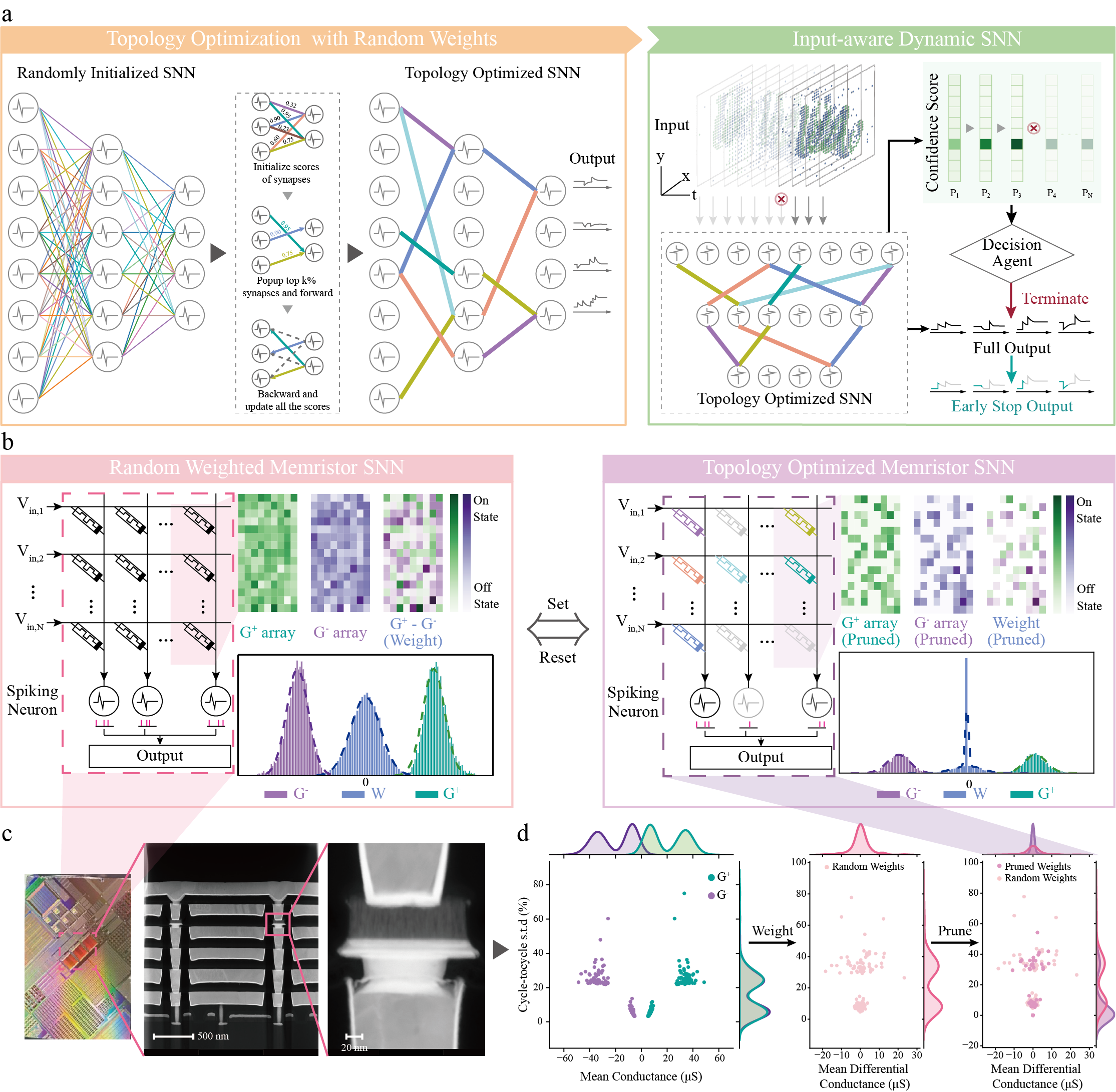}
\caption{\textbf{Overview of PRIME.}
\textbf{a,} The brain-inspired topology optimization of randomly initialized SNN (left). Initially, an overparameterized SNN with random memristor connections is generated using inherit programming stochasticity. Each random synaptic weight is then assigned a score \(s\), reflecting its importance. Synapses with the top \(k\%\) scores are retained, while others are pruned to form a subnet. The loss is calculated using this subnet and backpropagated through the entire network to optimize the scores. This process is iterated until convergence, yielding the optimal subnet. Further details are provided in \textbf{Methods}.  The brain-inspired input-aware dynamic SNN in inference (right). In inference, time-wise confidence is calculated, either as a softmax score or a consistency score, depending on the task. If the confidence meets the threshold policy, the inference terminates. Further details are provided in \textbf{Methods}.
\textbf{b,} The Schematic of Memristor-based SNN before (left) and after (right) topology optimization, consisting of memristor crossbar arrays and leaky integrate-and-fire neurons. Before pruning, the differential memristor pairs in the randomly overparameterized supernet (\(\textbf{G}^+\) and \(\textbf{G}^-\)) follows a mixture of Gaussian distributions. After pruning, the redundant memristor pairs are RESET, resulting in a conductance peak around zero.
\textbf{c,} Optical photo of the 40nm 256K memristor in-memory computing macro (left). Cross-sectional HAADF–STEM image of the memristor array (middle and right).
\textbf{d,} Joint distribution of the mean conductance and standard deviation of 128 randomly selected resistive memory cells in 10,000 reinstating programming cycles (left). Joint distribution of the 128 resistive differential pairs before (middle) and after pruning (right).}
\label{fig2}
\end{figure}



PRIME is systematically illustrated in Fig. ~\ref{fig2}.

Fig. ~\ref{fig2}a left panel illustrates the pruning optimization inspired by human brain's structural synaptic plasticity\cite{paolicelli2011synaptic, faust2021mechanisms}, which contributes to the maturation of human brain by preserving the functional synapses while eliminating redundant ones (Fig. ~\ref{fig1}b, left). In the training phase, random synaptic weights are generated by the inherent programming stochasticity of memristors (Fig. ~\ref{fig2}a, left). Each synapse has a pop-up score $s$, indicative of its significance. During the forward pass, the synapses with top $k\%$ pop-up scores are preserved while others are pruned. Subsequently, in the backward pass, these scores undergo gradient-based updates aimed at minimizing the training loss, thereby optimizing the topology of SNN (see \textbf{Methods}). This topology optimization gets rid of the expensive memristor conductance fine-tuning and transforms the memristor programming stochasticity into an advantage (see \textbf{Supplementary Note 1} for a more detailed theoretical proof of topology optimization for memristor-based SNN). 

In Fig. ~\ref{fig2}a right panel, drawing inspiration from the human brain's dynamic computational depth\cite{dynamic_com1, dynamic_com2, dynamic_com3, dynamic_com4, dynamic_com5}, which enhances the brain's ability to rapidly and efficiently recognize and adapt to new stimuli, we implement the input-aware dynamic early stop policy in the pruned memristive SNN during the inference phase. The input spikes to SNN are temporally spanned in a time window. As such, increase in SNN time steps increases the network performance, a common observation in computing with event data\cite{li2023input, li2024seenn, li2023unleashing}.  Here we introduce the confidence thresholding, where an agent computes the confidence scores of SNN outputs over different time steps. If the confidence score meets the predefined threshold, the inference process terminates (see \textbf{Methods} for details). This approach dynamically adjusts the computational load of each sample, aiming to reduce both latency and energy consumption, with minimal impact on the model's overall performance.

Hardware-wise, the memristor array is divided into positive and negative sub-arrays ($\textbf{G}^+$ and $\textbf{G}^-$), representing the weight matrix through the conductance difference (Fig. ~\ref{fig2}b, left). Initially, these memristors are insulating, leading to a weight distribution centered around zero conductance. After electroforming, these arrays exhibit random and analogue conductance, this translates to neural network weights following a mixture of two quasi-normal distributions due to the stochasticity in memristor programming. These random weights constitute the overparameterized supernet. During training, a subnet is sampled by prunning unnecessary connections, which is physically done by RESET the corresponding memristor pairs. After training, the conductance distribution in the optimized subnet shows a peak around zero conductance due to the pruned synapses (Fig. ~\ref{fig2}b, right). The integrated memristor arrays are the analogue cores of a hybrid analogue-digital computing system (Fig. ~\ref{fig2}c, \textbf{Supplementary Figure 1-2}), which implements vector-matrix multiplications. The digital core carries out non-matrix operations, including leaky integrate-and-fire neurons. The conductance distributions of selected memristors are presented in Fig. \ref{fig2}d, illustrating changes in the means and standard deviations of synaptic weights as pruning occurs.

\begin{figure}[!t]
\centering
\includegraphics[width=0.9\linewidth]{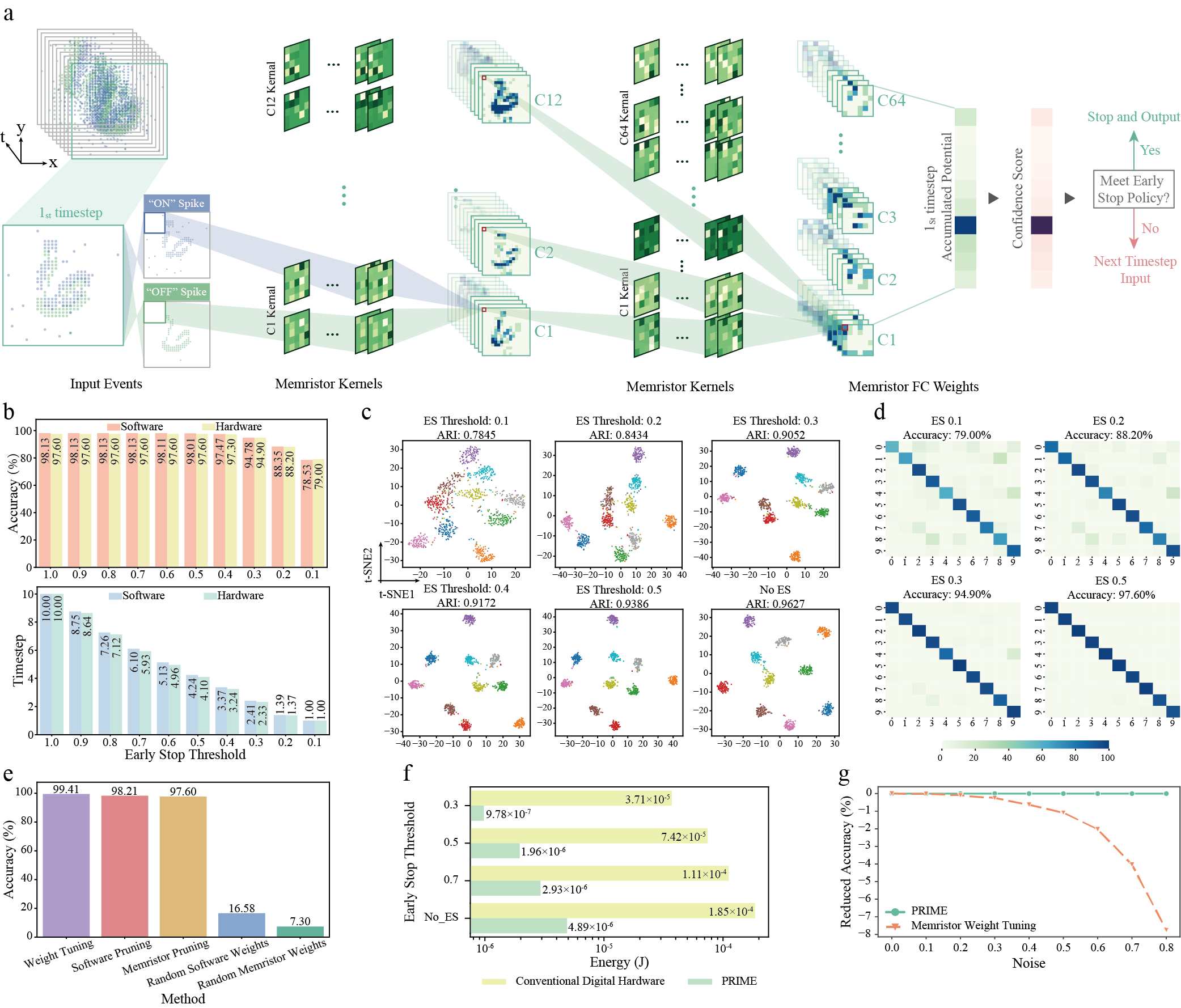}
\caption{\textbf{Experimental image classification for N-MNIST dataset with PRIME.}
\textbf{a,} Illusration of the convolutional SNN of PRIME during the inference, showing pruned random memristor kernels and associated feature maps in the N-MNIST classification. The network outputs at each time step are used to compute confidence scores.
\textbf{b,} The classification accuracy and dynamic latency (evaluated as the average timesteps of the test data) comparisons of hardware PRIME and software baseline at various early stop thresholds.
\textbf{c,} tSNE visualizations of feature maps from PRIME at different early stop thresholds, coloured according to ground truth labels.
\textbf{d,} Confusion matrices and classification accuracy of PRIME at different early stop thresholds.
\textbf{e,} The classification accuracy for SNNs by different optimization methods. \textbf{Weight Tuning}: Optimize the weights of software SNN through STBP. \textbf{Software Pruning}: Optimize the topology of software SNN,  with randomly initialized weights. \textbf{Memristor Pruning}: Optimize the topology of memristor-based SNN, where the random weights are produced by memristor programming stochasticity. \textbf{Random Software and Memristor Weights}: The SNNs are initialized with random weights, separately implemented in software and on memristors.
\textbf{f,} Comparison of the inference energy of a single image on a projected hybrid analogue-digital system and digital hardware at different early stop thresholds. The former shows a significant energy reduction due to in-memory computing.
\textbf{g,} Impact of memristor programming noise on accuracy between memristor-based SNNs optimized by different methods at various noise levels.
}
\label{fig3}
\end{figure}

\section*{Image classification for neuromorphic dataset using PRIME}
We first validate PRIME on classifying the representative N-MNIST dataset using a 3-layer spiking convolutional neural network (Fig. ~\ref{fig3}a). 

N-MNIST\cite{orchard2015converting} is a neuromorphic image datasets of 10 digits captured by dynamic vision sensors. Each sample, comprising `on' and `off' spike streams within a 34×34 pixel frame spanning 300ms, is processed into 10 time bins using SpikingJelly\cite{fang2023spikingjelly} (Fig. ~\ref{fig3}a). Here we use a supernet consisting of two convolution layers and a linear classification layer, with all synaptic weights initially mapped to random memristor conductance differentials. The softmax-based confidence score (Fig. ~\ref{fig3}a, see \textbf{Methods}) dynamically adjusts the inference timestep for each input sample, optimizing processing efficiency.

To assess PRIME's performance in terms of classification accuracy and inference efficiency, we first conduct a comparative analysis between the software baseline (Software) and PRIME (Hardware) across various early stop thresholds (Fig. ~\ref{fig3}b). The average timesteps in the test dataset are used as evaluation metric. PRIME closely matches the accuracy of the software baseline at timestep 10 for N-MNIST classification. Moreover, PRIME achieves significant latency reductions while maintaining high accuracy, thanks to the dynamic early stop using thresholding method. For example, at a threshold of 0.5, PRIME attains 97.60\% accuracy with about 59\% reduction in calculation cost (the average timesteps), and at a threshold of 0.3, it achieves 94.9\% accuracy with about a 77\% reduction in calculation cost. We then visualize the embedded features for the classification head under different early stop thresholds of PRIME (Fig. ~\ref{fig3}c, \textbf{Supplementary Figure 3a}) and software counterpart (\textbf{Supplementary Figure 3b}) using tSNE\cite{van2008visualizing}. The tSNE visualization results suggest that PRIME representations maintain a discernable subpopulation structure of ten clusters at the proper early stop threshold (Fig. ~\ref{fig3}c). The confusion matrices (Fig. ~\ref{fig3}d, \textbf{Supplementary Figure 4}) of PRIME is dominated by the diagonal elements at the proper threshold (e.g. 0.5) and hence corroborates the high classification accuracy with a substantial reduction in timesteps. 

We compare PRIME with other optimization methods (Fig. ~\ref{fig3}e). PRIME parallels the performance of weight-optimized SNN, while eliminating the need for weight fine-tuning. Additionally, compared to memristor-initialized SNNs (fixed random weights without any further optimization), PRIME shows significant performance improvement, consistent with software-pruned SNNs. This reveals that the memristor programming stochasticity is well-suited for generating random weights for topology optimization. The performance gap between software-pruning and software-initialized networks and that between memristor-pruning and memristor-initialized networks are similar, proving again that our proposed pruning algorithm is particularly effective for memristor-based networks.

Additionally, we present a comparative analysis of energy consumption for a single image classification between a projected hybrid analogue-digital system and state-of-the-art digital system including RTX4090 graphic processing unit (GPU) (Fig. ~\ref{fig3}f) (see \textbf{Supplementary Figure 5a}, \textbf{Supplementary Table 8} for comparison with other digital systems). The bottom panel (No early stop) illustrates that our PRIME can significantly decrease energy consumption, by approximately 37.83 times compared to the digital hardware, due to memristive in-memory computing. Moreover, the input-aware dynamic early stop further reduces energy consumption as the early stop threshold decreases (see \textbf{Supplementary Figure 6a} for detailed energy breakdown). This corroborates the superior energy efficiency of memristor in-memory computing.

We also compare PRIME with conventional memristor weights fine-tuning for network inference under different memristor programming noise. \cite{wang2019situ, qiangfei_1, li2018analogue}. For conventional weight fine-tuning, the inevitable programming stochasticity degrades the precision in mapping weights to memristor conductance, thus degrading the network inference performance as the noise increases. In contrast, PRIME leverages this programming stochasticity for weight initialization and its performance is not affected by the noise level.

We further evaluate PRIME using the DVS128 Gesture\cite{DVS128} dataset on spiking VGG-11 with simulated memristor conductance differentials (\textbf{Supplementary Figure 7}, \textbf{Methods}). The results indicate that PRIME achieves comparable accuracy with weight tuning and software pruning (\textbf{Supplementary Figure 7c}), while significantly reduces computational cost by 87.15\% with only a 1.74\% decrease in accuracy (\textbf{Supplementary Figure 7d-e}). Additionally, energy consumption is drastically reduced by 238.74 times compared to the RTX 4090 GPU (\textbf{Supplementary Figure 7f}). This simulated experiment shows that PRIME exhibits excellent scalability on larger memristor-based neural networks and promises substantial energy efficiency gains on deeper neural networks.

\begin{figure}[!t]
\centering
\includegraphics[width=0.9\linewidth]{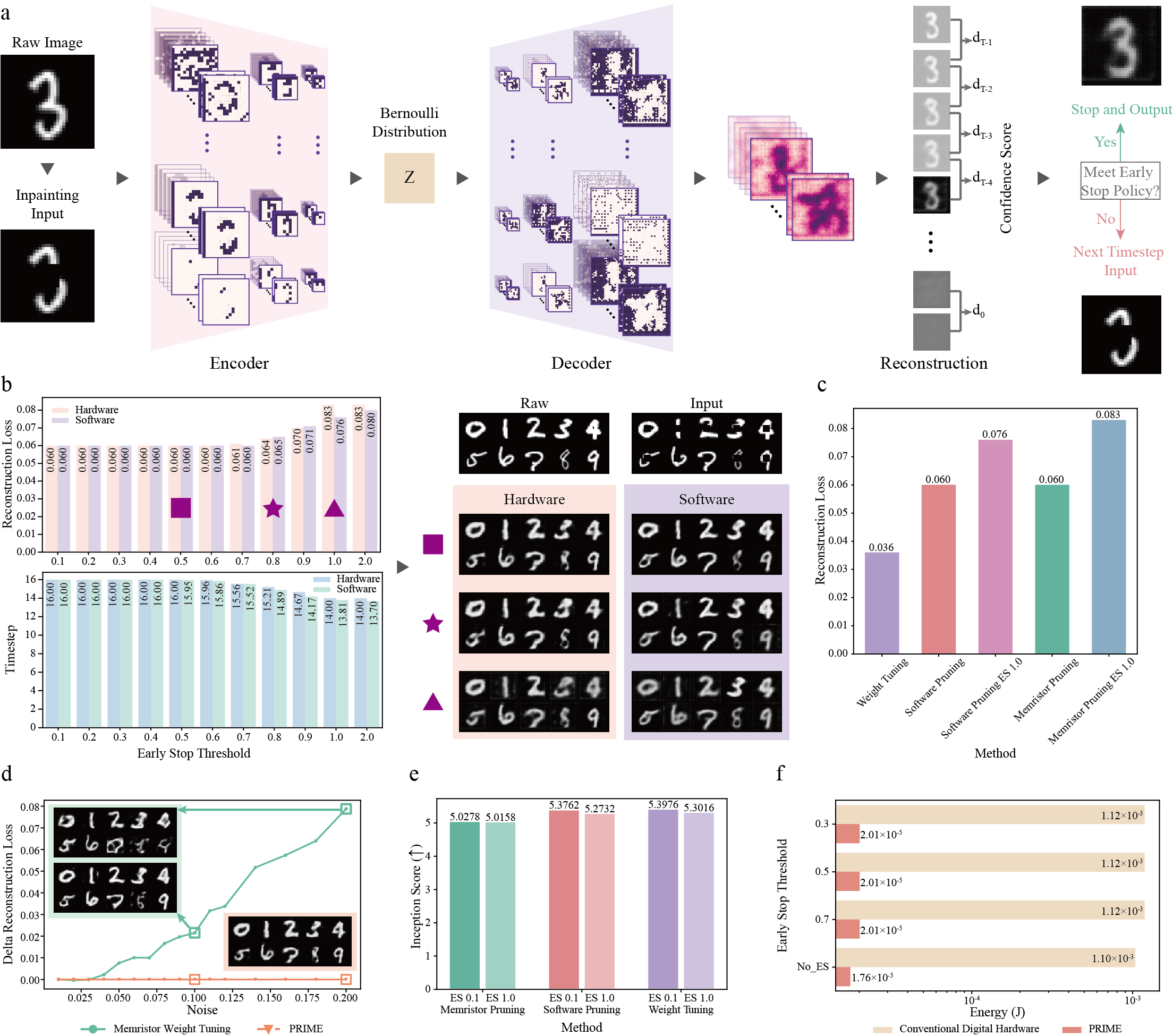}
\caption{\textbf{Experimental image inpainting of MNIST dataset with PRIME.} 
\textbf{a,} The spiking VAE of PRIME showing example feature maps during MNIST image inpainting. The latency of spiking VAE is dynamically adapted by the image consistence confidence score (\textbf{Methods}).
\textbf{b,} The reconstruction loss (\textbf{Methods}) and dynamic timestep (evaluated as the average timesteps of the test data) comparisons of hardware PRIME and software baseline at various early stop thresholds on MNIST (left). The raw, input, and reconstructed images at different thresholds (right). 
\textbf{c,} The reconstruction loss for SNNs optimized by different methods. \textbf{Weight Tuning}: Optimize the weights of software SNN through STBP. \textbf{Software Pruning and Software Pruning ES 1.0}: Optimize the topology of software SNN with randomly initialized weights. The dynamic early stop policy is either applied (former) or not applied (latter) in inference. \textbf{Memristor Pruning and Memristor Pruning ES 1.0}: Optimize the topology of memristor-based SNN, where the random weights are produced by memristor programming stochasticity. The dynamic early stop policy is either applied (former) or not applied (latter) in inference.
\textbf{d,} Impact of memristor programming noise on SNNs optimized by different methods at various noise levels.
\textbf{e,} The IS comparisons of SNNs by different optimization methods at various early stop thresholds.
\textbf{f,} Comparison of the inference energy of a projected hybrid analogue-digital system and digital hardware at different early stop thresholds.
}
\label{fig4}
\end{figure}

\section*{Image inpainting using PRIME}

Despite neuromorphic image classification, we extended our validation to more complex tasks such as image inpainting, utilizing a spiking variational autoencoder (spiking-VAE) \cite{kamata2022fully}. Image inpainting, the task of completing missing regions in images, typically employs generation models like VAEs due to their ability to come up with multiple perceptual outcomes\cite{yeh2017semantic, peng2021generating}. 

Here we use the representative MNIST\cite{lecun1998gradient} dataset. We purposely remove the central region of each MNIST image (Fig. ~\ref{fig4}a, see \textbf{Methods}). The PRIME employs a Spiking-VAE, incorporating encoder, decoder layers, and the Bernoulli sampling layer. The synaptic weights within encoder and decoder layers are first mapped to random memristor conductance differentials before pruning optimization. The confidence score measures the consistency of decoder output over consecutive frames, which dynamically regulates the inference timesteps for each image reconstruction, thereby enhancing energy efficiency and improving inference speed.

We first compare the experimental PRIME with software one across varying early stop thresholds (Fig. ~\ref{fig4}b). The results, displayed in the left panel of Fig. ~\ref{fig4}b, indicate that experimental PRIME shows consistent performance with software one in terms of reconstruction accuracy at different thresholds.  Notably, as the early stop threshold increases, there is an approximately 12.5\% reduction in latency with minimal impact on reconstruction quality. The experimentally reconstructed images at different thresholds are shown in the right panel of Fig. ~\ref{fig4}b, which can hardly be differentiated from those from MNIST dataset. 

We compare PRIME with other optimization methods (Fig. ~\ref{fig4}c). The pruning optimization shows a slightly increased reconstruction loss over weight optimization, while weight optimization is more notably limited by memristor programming noise (Fig. ~\ref{fig4}d) . In addition, early stop leads to a slightly higher reconstruction loss while saving inference computational load.  

The Inception Score\cite{barratt2018note} (IS) is a widely used metric for generative models. IS assesses the quality and diversity of generated images using a pre-trained classifier. A higher IS indicates that the generated images are both high-quality and diverse. The results illustrated in Fig. ~\ref{fig4}e demonstrate that PRIME is capable of achieving IS comparable to the weight optimized models on software.

Fig. ~\ref{fig4}f shows the estimated energy consumption for a single image reconstruction (see \textbf{Supplementary Figure 5b} and \textbf{Supplementary Figure 6b} for comparison with other digital systems). The findings indicate that PRIME significantly reduces energy consumption by a factor of 62.5, compared to SNNs implemented on mainstream digital hardware, which is particularly advantageous for edge AI.

We then evaluate PRIME using the Fashion-MNIST\cite{Fashion} dataset on larger neural networks with simulated memristor conductance differentials (\textbf{Supplementary Figure 8}, \textbf{Methods}). The results show that PRIME achieves comparable reconstruction accuracy and produces similar reconstructed images to weight-tuning models (\textbf{Supplementary Figure 8b}). Additionally, the early stop policy enables PRIME to reduce computational cost by 8.46\% with minimal loss (\textbf{Supplementary Figure 8c, d}). Moreover, energy consumption is significantly reduced by 102.82 times compared to the RTX 4090 GPU (\textbf{Supplementary Figure 8e}). This simulated experiment further demonstrates that PRIME has excellent scalability on larger memristor-based neural networks and can effectively process more complex datasets.

\begin{figure}[!t]
\centering
\includegraphics[width=0.9\linewidth]{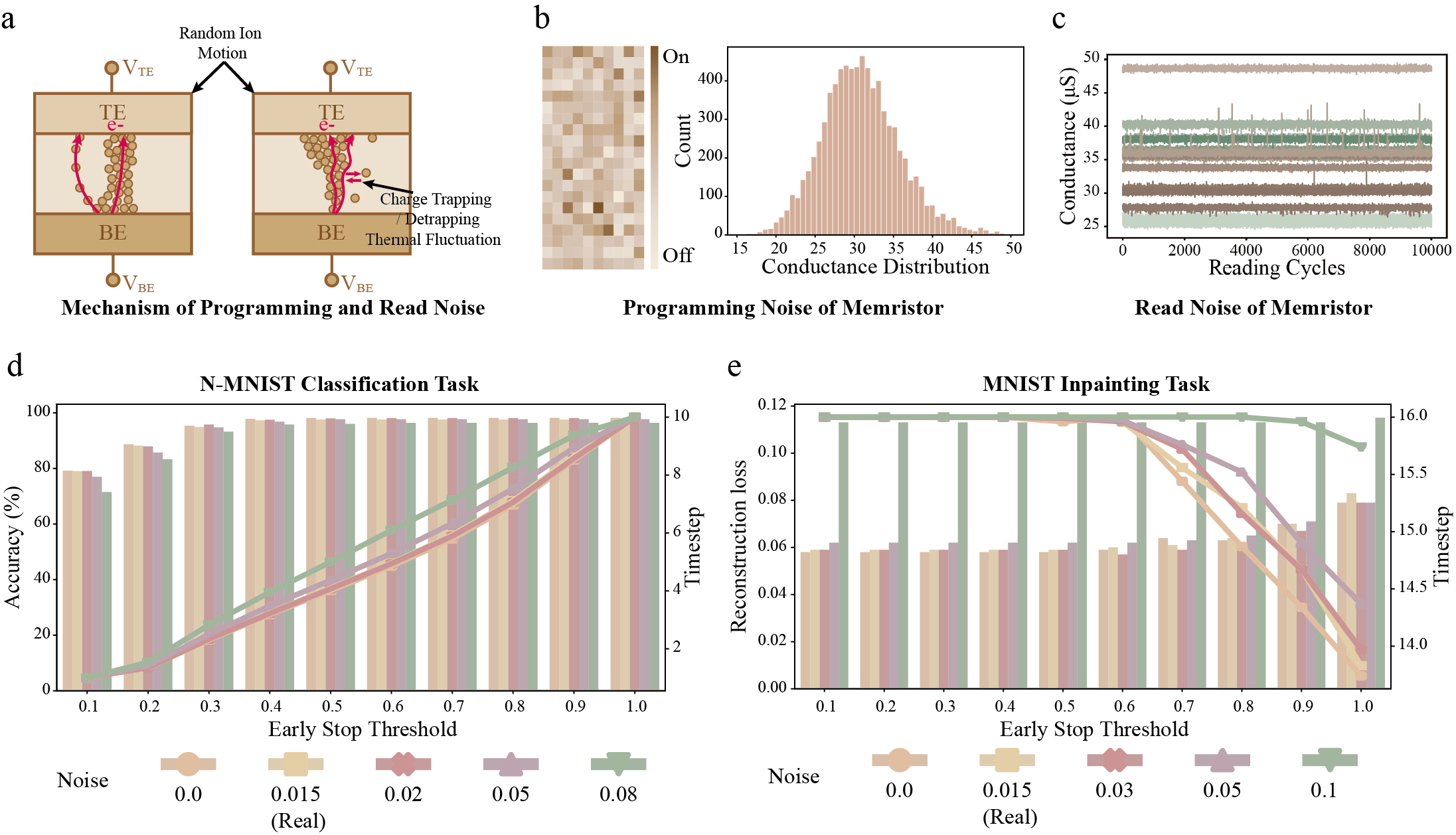}
\caption{\textbf{Memristor noise and impact on PRIME.} 
\textbf{a,} The mechanism of memristor programming and read noise. The random ion motion in filament formation and dissolution leads to programming noise, while charge trapping/detrapping and thermal fluctuation yields read noise.
\textbf{b,} The memristor programming noise represented as the heatmap and histgram, which follows a quasi-normal distribution for initializing random weights in PRIME.
\textbf{c,} The memristor read noise of 15 randomly selected memristors with 10,000 read cycles, showing clear conductance temporal fluctuation that degrades model's performance.
\textbf{d,} Noise robustness evaluation of PRIME at various early stop thresholds on N-MNIST classification.
\textbf{e,} Noise robustness evaluation of PRIME at various early stop thresholds on MNIST image inpainting.
}
\label{fig5}
\end{figure}

\section*{Impact of memristor read noise}
We further assess PRIME's capacity to mitigate memristor read noise using its input-aware dynamic early stop mechanism.

The memristive switching and transport mechanisms are illustrated in Fig. ~\ref{fig5}a. The charge transport through nanoscale conducting channels that are formed due to electrochemical reactions. This results in two types of noise: programming and read noise. The programming noise is due to random ionic motions in channel formation and rupture\cite{lin2020three}, giving rise to cycle-to-cycle and device-to-device variation in programming conductance  (Fig. ~\ref{fig5}b), which can be effectively addressed by our proposed pruning optimization. The read noise is the temporal fluctuation of memristor conductance due to charge trapping and detrapping (e.g. random telegraphic noise) and thermal noise \cite{sebastian2019computational, sebastian2020memory}. (Fig. ~\ref{fig5}c).

To demonstrate PRIME's effectiveness in mitigating memristor read noise, we experiment on various early stop thresholds with different levels of simulated Gaussian read noise (Fig. ~\ref{fig5}d-e, see \textbf{Supplementary Note 2} for the theoretical analysis of the impact of read noise on PRIME, and \textbf{Supplementary Figure 9} for simulated conductance fluctuations). PRIME demonstrates remarkable stability in terms of model performance within the typical memristor read noise range of 0.01 to 0.03 (see definition of noise scale in \textbf{Methods}). In high-noise scenarios (e.g., 0.08, 0.1), PRIME shows more performance degradation in inpainting (Fig. ~\ref{fig5}e), while the impact on classification is relatively minor (Fig. ~\ref{fig5}d).  Additionally, the study shows that variations in early stop thresholds affect timesteps and accuracy, especially under high-noise conditions, because  confidence thresholding of PRIME dynamically balance timesteps and network performance.

\section*{Discussion}
Although ANNs have shown unprecedented development, they still cannot parallel the brain in terms of adaptability and energy efficiency. 

PRIME offers a potential solution via the hardware-software co-design. In terms of hardware, the memristive neuromorphic computer mimics the in-memory computing of the brain. In addition, it also practises pruning optimization that is inspired by biological structural plasticity, making it robust to both programming and read noise of memristors. In terms of software, inspired by brain's dynamic adjustment of computational depth, we introduce an input-aware dynamic early stop policy for SNN using confidence score, which further boosts energy efficiency and inference speed.

Evaluations on neuromorphic datasets reveals that PRIME matches the accuracy of static weight-optimized SNNs in software, while exhibiting significant advantages in energy consumption and inference speed. PRIME achieves 37.83\(\times\) energy efficiency improvements and 77.0\% computational load savings in neuromorphic image classification. This performance motivates the application of PRIME to complex tasks like image inpainting using a spiking VAE. PRIME shows significantly improved energy efficiency in these tasks, with 62.50\(\times\) reduction in energy consumption compared to baseline SNNs and 12.5\% computational load savings with minimal reduction in performance. This is corroborated by the similar IS achieved by PRIME and the software baseline. And PRIME further exhibits excellent scalability on larger and deeper neural networks. Additionally, PRIME effectively mitigates memristor programming and read noise, a notable challenge in emerging neuromorphic hardware.

In conclusion, PRIME offers a precise, energy-efficient, and low-latency framework for future neuromorphic computing.

\section*{Methods}
\subsection*{Fabrication of Resistive Memory Chip}

Utilizing the advanced 40nm technology node, the engineered resistive memory device showcases a sophisticated 512×512 crossbar configuration. This intricate arrangement features resistive memory elements strategically positioned between 4 and 5 metal layers, employing a backend-of-line fabrication technique. Each cell within this array is meticulously crafted, consisting of bottom and top electrodes (BE and TE), complemented by a dielectric layer made of transition-metal oxide. The fabrication process begins with the precise patterning of the BE via, which boasts a 60nm diameter, achieved through photolithography and subsequent etching. This is followed by the deposition of TaN using physical vapor deposition techniques, capped with a 10nm TaN buffer layer for enhanced stability. A thin, 5nm layer of Ta is then applied and subjected to oxidation, culminating in the formation of an 8nm-thick TaO$_x$ dielectric layer. The construction of the TE involves a carefully sequenced deposition of 3nm Ta and 40nm TiN, also via physical vapor deposition. The fabrication process is finalized with the deposition of the requisite interconnection metals, adhering to conventional logic processing protocols. Within this architecture, cells aligned in the same row are interconnected through BE, whereas column-aligned cells share TE connectivity. A post-fabrication annealing step, conducted for 30 minutes at 400°C under vacuum conditions, significantly enhances the chip's performance. This meticulous manufacturing process yields a resistive memory chip with exemplary performance metrics, characterized by its high operational yield and exceptional endurance capabilities.

\subsection*{Hybrid Analogue–Digital Hardware System}

The innovative hybrid system merges analogue-digital technologies, integrating a 40nm resistive memory chip with a Xilinx ZYNQ system-on-chip (SoC). This SoC amalgamates a field-programmable gate array (FPGA) with an advanced RISC machines (ARM) processor, all mounted on a printed circuit board (PCB) for cohesive operation. The resistive memory chip is designed to function in three distinct modes, each crucial for the edge pruning topology optimization: electroform mode, reset mode, and multiplication mode.

In the electroform mode, a controlled dielectric breakdown is initiated within the resistive memory arrays, effectively creating random conductance matrices. This is achieved by biasing all source lines (SLs) to a predetermined programming voltage, delivered by an eight-channel digital-to-analogue converter (DAC, DAC80508 from Texas Instruments) boasting a 16-bit resolution. Meanwhile, bit lines (BLs) are grounded, and word lines (WLs) receive biasing from the DAC to enforce a compliance current across the cells, thereby averting a hard breakdown. The nuances of the SL voltage amplitude and duration are key to shaping the post-breakdown conductance distribution and its sparsity.

Transitioning to the reset mode enables the reversion of a resistive memory cell to its non-conductive state, wherein a selected BL is biased via the DAC, the corresponding SL is grounded, and the remaining SLs are left in a floating state. The multiplication mode involves the utilization of a 4-channel analogue multiplexer (CD4051B, Texas Instruments) coupled with an 8-bit shift register (SN74HC595, Texas Instruments), which collectively apply a DC voltage across the BLs of the resistive memory chip.

Throughout each phase of training, the chip undergoes readings, and the resultant multiplication values, manifested as currents from the SLs, are transformed into voltages. This conversion is facilitated by trans-impedance amplifiers (OPA4322-Q1, Texas Instruments) and analogue-to-digital converters (ADS8324, Texas Instruments, with a 14-bit resolution), with the processed data subsequently relayed to the Xilinx SoC for advanced computation. The FPGA component of the SoC is intricately designed to manage the resistive memory operations and facilitate data exchange with the ARM processor via a direct memory access control unit, employing double-data rate memory access. Furthermore, the FPGA is tasked with the hardware implementation of certain neural network functions, including activation and pooling, enhancing the system's overall computational efficacy.

\subsection*{Reset and Set Operations}

\subsubsection*{Edge Pruning Topology Optimization}

The process of synaptic pruning within this system is physically manifested by transitioning the relevant differential pairs of resistive memory into an off-state via the reset operation. Conversely, the reactivation of these synaptical weights is facilitated by restoring the resistive memory cells to their conductive states through a set operation. This set operation is executed by administering uniform pulses, characterized by a 3.3V amplitude and a 300ns duration, to the bit line of the resistive memory array. This procedure revives the previously pruned cells, reintegrating them into the active sub-network. On the other hand, the reset operation is effected through the application of uniform pulses, this time with a 2.6V amplitude and a 400ns duration, to the source line of the resistive memory array, effectively eliminating the conductive pathway.

It is crucial to highlight the substantial distinction between the off-state and the conducting state of these cells. This differentiation underscores the fact that programming the cells to precise conductance values is not requisite for the system's functionality, allowing for a more flexible and efficient approach to the modulation of conductive states within the network.

\subsection*{PRIME model}
\subsubsection*{Spiking Neuron Model}
The iterative leaky integrate-and-fire (iLIF) spiking model\cite{wu2019direct} is utilized, which is a LIF model solved using the Euler method.
\begin{equation}
\begin{aligned}
    u_t &= \tau_{decay}u_{t-1}(1-o_{t-1})+x_t \\
    o_t &= H(u_t-V_{th})
\label{eq:Dynamics equations of LIF},
\end{aligned}
\end{equation}
where $\tau_{decay}$ represents membrane decay, $u_t$, $u_{t-1}$, $o_t$, $o_{t-1}$ are the membrane potential and spike output (i.e. 0 or 1) at time step $t$ and $t-1$. $x_t$ denotes the weighted sum of spikes from the connected neurons, where $x_{j,t} = \sum_{j} w_{j}o_{j,t}$. $H(x)$ represents the heaviside step function which will generate a spike when $x>0$. $V_{th}$ is the threshold potential. Due to the discontinuity of the heaviside step function we utilize the approach of pseudo-derivative to solve the issue. In detail, we approximate it as follows:
\begin{equation}
\frac{\partial{o_t}}{\partial{u_t}} = \frac{1}{a} sign(|u_t-V_{th}|<\frac{a}{2})
\label{eq:Pseudo-derivative function},
\end{equation}
where $a$ is a hyperparameter defined as 1 within the context of this study.

\subsubsection*{Pruning Topology Optimization with random weights}
In the pruning topology optimization method, there are two sets of parameters, i.e. randomly distributed weights $W$, and pop-up scores $S$. Initially, a spiking neural network with random weights is established using an analogue resistive memory chip. Unlike the traditional co-design model, where the weights $W$ are expensively tuned, this configuration involves learning a pop-up score $s$ for each synaptic weight $w$, while maintaining the weights in their initial random distribution.\\
The pruning process can be divided into two phases, the forward pass and the backward pass. On the forward pass, the hardware synapses are selected with top-k\% highest pop-up scores in each layer according to the predefined sparsity, leading to a pop-up score based sub-network. 
Inputs are then fed into the sub-network for forward propagation and loss evaluation. The input of neuron \textit{i} in layer \textit{l} can be computed as:
\begin{equation}
I_\textit{i} = {\sum_{j\in{V^{l-1}}}w_{ij}Z_jH(s_{ij})}
\label{eq:forward when pruning},
\end{equation}
where the $j$ neuron in layer $l-1$ is connected with the $i$ neuron via the synaptic weight $w_{ij}$. $Z_j$ denotes the output of the neuron $j$. And the $H(s_{ij})=1$ if $s_{ij}$ is in the top-k\% pop-up scores in layer $l$. On the backward pass, the general digital processor calculates the loss function's gradients to optimize the scores of all synapses with the random weights fixed. The pop-up score is updated using the straight-through estimator~\cite{bengio2013estimating,ramanujan2020s}:
\begin{equation}
{s}_{ij} \gets {s_{ij} - \alpha \frac{\partial{L}}{\partial{I_i}} w_{ij}Z_j}
\label{eq:straight-through gradient},
\end{equation}
where $s_{ij}$, $\alpha$, $\frac{\partial{L}}{\partial{I_i}}$ and $w_{ij}Z_j$ denote the pop-up score between the connected neuron $i$ and $j$, the learning rate, the partial derivative of loss ($L$) with respect to the input of node $i$ and the weighted output of node $j$, respectively. These processes are repeated until a well-performed sub-network is selected from the randomly initialized neural network.

\subsubsection*{Dynamic confidence thresholding method}

The dynamic confidence thresholding method is input-dependent, which dynamically decides which time step to early stop for each input sample during inference, leading to faster, lower overhead, more energy-efficient edge computing. Consequently, the number of time steps varies for each input, potentially allowing for a reduction in the average time steps required per inference.

In the dynamic confidence thresholding method, the confidence metric is calculated based on the output, and then utilized to determine the optimal point to prematurely conclude the inference process. The confidence calculation varies across different tasks. Considering the classification tasks, given an input $X$ with a label $Y$, the prediction probability distribution for each time step $t$ is usually calculated by Softmax, $\textbf{P}(\overline{Y} = Y) = softmax(f(X_t))=[p_1, p_2, ..., p_N]$, where $\overline{Y}$ is the predicted output label, N is the number of categories. Then, the confidence is defined as $\overline{P_t} = max(\textbf{P})$,
\begin{equation}
\overline{P_t} = max(\frac{e^{f(X_t)_i / \alpha}}{\sum_{n=1}^{N}e^{f(X_t)_i / \alpha}}) \quad i=1,2,...,N
\label{eq:Confidence calculation for classification task},
\end{equation}
where $\overline{P_t}$ represents the calculated confidence of time step $t$, $\alpha$ denotes denotes a scale parameter designed to prevent the saturation of confidence levels. Given a predefined threshold $\beta_1$, the inference process will terminate early, using the output at this time step for prediction, if the confidence reaches or exceeds $\beta_1$ (i.e. $\overline{P_t} \geq \beta_1$).

Considering the impainting task, for a given impainting input image $X$, the output at each time step is the reconstructed image $\overline{X_t}$. The confidence for this task is determined by the consistence calculation:
\begin{equation}
\overline{P_t} = \Vert X_{t} - X_{t-1} \Vert
\label{eq:Confidence calculation for imapinting task},
\end{equation}
where $\Vert \cdot \Vert$ represents the $L1$ norm, which is utilized to quantify the difference between two images at consecutive two time steps. When the predefined threshold $\beta_2$ is defined, the inference process will terminate early when the condition $\overline{P_{t}} < \overline{P_{t-1}} < \beta_2$ is met. This implies that when the change in consecutively generated images becomes minimal, indicating a negligible difference, the process is considered to have reached its termination point.

\subsection*{Details of the experiments}
\subsubsection*{Classification on neuromorphic image dataset}
The event-based dataset, N-MNIST\cite{orchard2015converting} is utlized to evaluate the performance of our model. The N-MNIST dataset is a spiking version of the MNIST dataset, created using a Dynamic Vision Sensor (DVS) mounted on a pan-tilt unit. It comprises 60,000 training samples and 10,000 testing samples, distributed across 10 classes representing digits from '0' to '9'. Each sample in this dataset includes 'on' and 'off' spike events and is represented in a resolution of 34×34 pixels, spanning a duration of 300ms. As shown in Fig. \ref{fig3}a, the 12C5-P2-64C5-P2 architecture is configured. The detailed training settings and default parameters of PRIME on this task is shown in \textbf{Supplementary Table 2} and \textbf{Supplementary Table 5}, respectively.

The event-based dataset, DVS128 Gesture\cite{DVS128} is utlized to evaluate the scalability of our model. The DVS128 Gesture dataset is a neuromorphic dataset designed for gesture recognition tasks using DVS.  The dataset consists of 11 different human actions, including hand clapping, hand waving, and arm rotation, among others. Each sample in this dataset includes 'on' and 'off' spike events and is represented in a resolution of 128×128 pixels. As shown in \textbf{Supplementary Figure 7}, the spiking VGG-11 architecture is configured. The detailed training settings and default parameters of PRIME on this task is shown in \textbf{Supplementary Table 3} and \textbf{Supplementary Table 6}, respectively.

\subsubsection*{Evaluation metric for classification}
\emph{Accuracy}. $\text{Accuracy} = \frac{\sum_{i} 1(y_i = \hat{y}_i)}{n}$, where $y_i$ represents the predicted output for sample $i$, $\hat{y}_i$ denotes the ground truth, and $n$ is the total number of samples.\vspace{4pt}

\noindent \emph{Average time steps}. $\text{Average ts} = \frac{\sum_{i} ts_i}{n}$, where $ts_i$ denotes the inference time steps for the sample $i$, and the $n$ represents the total number of samples. \vspace{4pt}

\noindent \emph{ARI}. $\text{ARI} = \frac{ \sum_{ij} \binom{n_{ij}}{2} - \left[ \sum_i \binom{a_i}{2} \sum_j \binom{b_j}{2} \right] / \binom{N}{2} }{ \frac{1}{2} \left[ \sum_i \binom{a_i}{2} + \sum_j \binom{b_j}{2} \right] - \left[ \sum_i \binom{a_i}{2} \sum_j \binom{b_j}{2} \right] / \binom{N}{2} }$, where $n_{ij}$ represents the number of elements in both cluster $i$ and cluster cluster $j$, $a_i$ and $b_j$ is the sum of elements in row $i$ and column $j$, indicating the total elements in cluster $i$ and $j$, $N$ is the totam number of elements, $\binom{n}{2}$ denotes the binomial coefficient, representing the number of unique pairs that can be formed from $n$ items. The Adjusted Rand Index (ARI) is a measure used to evaluate the similarity between two data clusterings. A higher ARI value, closer to 1, indicates greater similarity between the two clusterings, suggesting a more accurate clustering process.\vspace{4pt}

\noindent \emph{Energy Consumption}. The detailed components of energy consumption and their calculation methods are illustrated in \textbf{Supplementary Table 6}.

\subsubsection*{Inpainting on image dataset}
The MNIST\cite{lecun1998gradient} dataset is comprised of single-channel grayscale images, representing digits from '0' to '9', thus encompassing 10 distinct classes. Each image in this dataset is characterized by a resolution of 28×28 pixels. The dataset is divided into two subsets: 60,000 images for training and 10,000 for testing. In the inpainting task, as illustrated in Fig. \ref{fig4}a, the central 8x8 pixel region of each image is obscured. These modified images serve as inputs to a spiking Variational Autoencoder\cite{kamata2022fully} (spiking-VAE), which aims to reconstruct the original, unaltered image (the ground truth). The network's configuration follows the same structure as outlined in Kamata's work. However, a notable modification is made in the number of channels in both the encoder and decoder components of the network; they are adjusted to 32 channels, a reduction from the originally specified 64 and 128 channels. The detailed training settings and default parameters of PRIME on image inpainting is presented in \textbf{Supplementary Table 4} and \textbf{Supplementary Table 7}, respectively.

Fashion-MNIST\cite{Fashion} consists of 70,000 grayscale images, each with a resolution of 28 × 28 pixels. The images represent 10 different categories of fashion items, such as T-shirts, trousers, pullovers, dresses, coats, sandals, shirts, sneakers, bags, and ankle boots. Each category contains 7,000 images, with 60,000 images in the training set and 10,000 images in the test set. As illustrated in \textbf{Supplementary Figure 8}, the network's configuration follows deeper and larger structures. The detailed training settings and default parameters of the experiment is presented in \textbf{Supplementary Table 4} and \textbf{Supplementary Table 7}, respectively.

\subsubsection*{Evaluation metric for inpainting}
\emph{Reconstruction loss}. $L_{reconstruction} = MSE(x, \hat{x})$, where $x$ and $\hat{x}$ distinctly represent the ground truth and reconstructed images, respectively.\vspace{4pt}

\noindent \emph{Inception Score}. $IS = \exp\left(\mathbb{E}_{\mathbf{x}\sim p_g}\left[D_{KL}( p(y|\mathbf{x}) || p(y) )\right]\right)$, where $\mathbf{x}$ is the generated data, $p_g$ is the model's generated data distribution, $p(y)$  is the marginal class distribution over the generated data, $p(y|\mathbf{x})$ represents the conditional class distribution given the generated sample $\mathbf{x}$, typically obtained from an Inception model. The Inception Score (IS) is commonly used to evaluate the quality of images generated by models like VAEs. The Inception Score measures the diversity and quality of the generated images, where a higher score generally indicates better image quality and more variety in the generated samples.\vspace{4pt}

\noindent \emph{Fréchet Inception Distance}. FID is a metric commonly used to assess the quality of generated images in generative models, similar to the Inception Score (IS). FID quantifies the similarity between the feature representations of real and generated images using a pre-trained Inception network. It is calculated as $
FID = \|\mathbf{\mu}_r - \mathbf{\mu}_g\|^2 + \text{Tr}(\Sigma_r + {\Sigma}_g - 2({\Sigma}_r{\Sigma}_g)^{1/2})$, where \( \mathbf{\mu}_r \) and \( \mathbf{\mu}_g \) are the mean feature vectors of the real and generated images, respectively, and \( {\Sigma}_r \) and \({\Sigma}_g \) are their covariance matrices. FID measures both the quality and diversity of the generated images, with lower values indicating better performance. It captures the perceptual similarity between the real and generated image distributions, providing a comprehensive assessment of the generative model's performance. The FID results are depicted in \textbf{Supplementary Table 1}.\vspace{4pt}

\noindent \emph{Energy Consumption}. The detailed components of energy consumption and their calculation methods are illustrated in \textbf{Supplementary Table 6}.

\subsubsection*{Memristor read conductance simulation}
The memristor conductance with different levels of read noise is simulated using Gaussian noise, which is formulated as 
\begin{equation}
mem\_g = mem\_g + A \times \text{noise\_scale} \times mem\_g
\label{eq: Noise Simulation},
\end{equation}
where the $mem\_g$ denotes the memristor conductance, $A$ is a random variable following a normal distribution with mean 0 and variance 1, and $noise\_scale$ represents the level of read noise in the memristor.

\section*{Acknowledgements}
This research is supported by the National Key R\&D Program of China (Grant No. 2022YFB3608300), the National Natural Science Foundation of China (Grant Nos. 62122004, 62374181, 61888102, 61821091), the Strategic Priority Research Program of the Chinese Academy of Sciences (Grant No. XDB44000000), Beijing Natural Science Foundation (Grant No. Z210006), and Hong Kong Research Grant Council (Grant Nos. 27206321, 17205922, 17212923). This research is also partially supported by ACCESS – AI Chip Center for Emerging Smart Systems, sponsored by Innovation and Technology Fund (ITF), Hong Kong SAR.

\section*{Competing Interests}
The authors declare no competing interests.

\section*{Data availability}
The N-MNIST\cite{orchard2015converting}, DVS128 Gesture\cite{DVS128}, MNIST\cite{lecun1998gradient}, and Fashion-MNIST\cite{Fashion} are publicly available. All other measured data are freely available upon reasonable request. 

\section*{Code availability}
Source codes in Pytorch are also available at Github: 
 \url{https://github.com/bo-wang-up/PRIME}. 
\bibliography{reference}

\begin{thebibliography}{10}
\urlstyle{rm}
\expandafter\ifx\csname url\endcsname\relax
  \def\url#1{\texttt{#1}}\fi
\expandafter\ifx\csname urlprefix\endcsname\relax\def\urlprefix{URL }\fi
\expandafter\ifx\csname doiprefix\endcsname\relax\def\doiprefix{DOI: }\fi
\providecommand{\bibinfo}[2]{#2}
\providecommand{\eprint}[2][]{\url{#2}}

\bibitem{lecun2015deep}
\bibinfo{author}{LeCun, Y.}, \bibinfo{author}{Bengio, Y.} \& \bibinfo{author}{Hinton, G.}
\newblock \bibinfo{journal}{\bibinfo{title}{Deep learning}}.
\newblock {\emph{\JournalTitle{nature}}} \textbf{\bibinfo{volume}{521}}, \bibinfo{pages}{436--444} (\bibinfo{year}{2015}).

\bibitem{philips_1}
\bibinfo{author}{Kuzum, D.}, \bibinfo{author}{Jeyasingh, R.~G.}, \bibinfo{author}{Lee, B.} \& \bibinfo{author}{Wong, H.-S.~P.}
\newblock \bibinfo{journal}{\bibinfo{title}{Nanoelectronic programmable synapses based on phase change materials for brain-inspired computing}}.
\newblock {\emph{\JournalTitle{Nano letters}}} \textbf{\bibinfo{volume}{12}}, \bibinfo{pages}{2179--2186} (\bibinfo{year}{2012}).

\bibitem{yuchao_1}
\bibinfo{author}{Zhang, T.} \emph{et~al.}
\newblock \bibinfo{journal}{\bibinfo{title}{Memristive devices and networks for brain-inspired computing}}.
\newblock {\emph{\JournalTitle{physica status solidi (RRL)--Rapid Research Letters}}} \textbf{\bibinfo{volume}{13}}, \bibinfo{pages}{1900029} (\bibinfo{year}{2019}).

\bibitem{chang2023survey}
\bibinfo{author}{Chang, Y.} \emph{et~al.}
\newblock \bibinfo{journal}{\bibinfo{title}{A survey on evaluation of large language models}}.
\newblock {\emph{\JournalTitle{ACM Transactions on Intelligent Systems and Technology}}}  (\bibinfo{year}{2023}).

\bibitem{zhao2023survey}
\bibinfo{author}{Zhao, W.~X.} \emph{et~al.}
\newblock \bibinfo{journal}{\bibinfo{title}{A survey of large language models}}.
\newblock {\emph{\JournalTitle{arXiv preprint arXiv:2303.18223}}}  (\bibinfo{year}{2023}).

\bibitem{videoworldsimulators2024}
\bibinfo{author}{Brooks, T.} \emph{et~al.}
\newblock \bibinfo{journal}{\bibinfo{title}{Video generation models as world simulators}}.
\newblock {\emph{\JournalTitle{Technical Report}}}  (\bibinfo{year}{2024}).

\bibitem{yuchao_2}
\bibinfo{author}{Kumar, S.}, \bibinfo{author}{Wang, X.}, \bibinfo{author}{Strachan, J.~P.}, \bibinfo{author}{Yang, Y.} \& \bibinfo{author}{Lu, W.~D.}
\newblock \bibinfo{journal}{\bibinfo{title}{Dynamical memristors for higher-complexity neuromorphic computing}}.
\newblock {\emph{\JournalTitle{Nature Reviews Materials}}} \textbf{\bibinfo{volume}{7}}, \bibinfo{pages}{575--591} (\bibinfo{year}{2022}).

\bibitem{JJYang_1}
\bibinfo{author}{Yang, J.~J.}, \bibinfo{author}{Strukov, D.~B.} \& \bibinfo{author}{Stewart, D.~R.}
\newblock \bibinfo{journal}{\bibinfo{title}{Memristive devices for computing}}.
\newblock {\emph{\JournalTitle{Nature nanotechnology}}} \textbf{\bibinfo{volume}{8}}, \bibinfo{pages}{13--24} (\bibinfo{year}{2013}).

\bibitem{philips_2}
\bibinfo{author}{Ielmini, D.} \& \bibinfo{author}{Wong, H.-S.~P.}
\newblock \bibinfo{journal}{\bibinfo{title}{In-memory computing with resistive switching devices}}.
\newblock {\emph{\JournalTitle{Nature electronics}}} \textbf{\bibinfo{volume}{1}}, \bibinfo{pages}{333--343} (\bibinfo{year}{2018}).

\bibitem{huaqiang_1}
\bibinfo{author}{Ning, H.} \emph{et~al.}
\newblock \bibinfo{journal}{\bibinfo{title}{An in-memory computing architecture based on a duplex two-dimensional material structure for in situ machine learning}}.
\newblock {\emph{\JournalTitle{Nature nanotechnology}}} \textbf{\bibinfo{volume}{18}}, \bibinfo{pages}{493--500} (\bibinfo{year}{2023}).

\bibitem{luwei_1}
\bibinfo{author}{Zidan, M.~A.}, \bibinfo{author}{Strachan, J.~P.} \& \bibinfo{author}{Lu, W.~D.}
\newblock \bibinfo{journal}{\bibinfo{title}{The future of electronics based on memristive systems}}.
\newblock {\emph{\JournalTitle{Nature electronics}}} \textbf{\bibinfo{volume}{1}}, \bibinfo{pages}{22--29} (\bibinfo{year}{2018}).

\bibitem{dynamic_com1}
\bibinfo{author}{Lennie, P.}
\newblock \bibinfo{journal}{\bibinfo{title}{The cost of cortical computation}}.
\newblock {\emph{\JournalTitle{Current biology}}} \textbf{\bibinfo{volume}{13}}, \bibinfo{pages}{493--497} (\bibinfo{year}{2003}).

\bibitem{dynamic_com2}
\bibinfo{author}{Mattar, M.~G.} \& \bibinfo{author}{Lengyel, M.}
\newblock \bibinfo{journal}{\bibinfo{title}{Planning in the brain}}.
\newblock {\emph{\JournalTitle{Neuron}}} \textbf{\bibinfo{volume}{110}}, \bibinfo{pages}{914--934} (\bibinfo{year}{2022}).

\bibitem{dynamic_com3}
\bibinfo{author}{Snider, J.}, \bibinfo{author}{Lee, D.}, \bibinfo{author}{Poizner, H.} \& \bibinfo{author}{Gepshtein, S.}
\newblock \bibinfo{journal}{\bibinfo{title}{Prospective optimization with limited resources}}.
\newblock {\emph{\JournalTitle{PLoS computational biology}}} \textbf{\bibinfo{volume}{11}}, \bibinfo{pages}{e1004501} (\bibinfo{year}{2015}).

\bibitem{dynamic_com4}
\bibinfo{author}{Keramati, M.}, \bibinfo{author}{Smittenaar, P.}, \bibinfo{author}{Dolan, R.~J.} \& \bibinfo{author}{Dayan, P.}
\newblock \bibinfo{journal}{\bibinfo{title}{Adaptive integration of habits into depth-limited planning defines a habitual-goal--directed spectrum}}.
\newblock {\emph{\JournalTitle{Proceedings of the National Academy of Sciences}}} \textbf{\bibinfo{volume}{113}}, \bibinfo{pages}{12868--12873} (\bibinfo{year}{2016}).

\bibitem{dynamic_com5}
\bibinfo{author}{Van~Opheusden, B.}, \bibinfo{author}{Galbiati, G.}, \bibinfo{author}{Bnaya, Z.}, \bibinfo{author}{Li, Y.} \& \bibinfo{author}{Ma, W.~J.}
\newblock \bibinfo{title}{A computational model for decision tree search.}
\newblock In \emph{\bibinfo{booktitle}{CogSci}} (\bibinfo{year}{2017}).

\bibitem{llinas1982electrophysiology}
\bibinfo{author}{Llin{\'a}s, R.} \& \bibinfo{author}{Jahnsen, H.}
\newblock \bibinfo{journal}{\bibinfo{title}{Electrophysiology of mammalian thalamic neurones in vitro}}.
\newblock {\emph{\JournalTitle{Nature}}} \textbf{\bibinfo{volume}{297}}, \bibinfo{pages}{406--408} (\bibinfo{year}{1982}).

\bibitem{hahn2019portraits}
\bibinfo{author}{Hahn, G.}, \bibinfo{author}{Ponce-Alvarez, A.}, \bibinfo{author}{Deco, G.}, \bibinfo{author}{Aertsen, A.} \& \bibinfo{author}{Kumar, A.}
\newblock \bibinfo{journal}{\bibinfo{title}{Portraits of communication in neuronal networks}}.
\newblock {\emph{\JournalTitle{Nature Reviews Neuroscience}}} \textbf{\bibinfo{volume}{20}}, \bibinfo{pages}{117--127} (\bibinfo{year}{2019}).

\bibitem{wozniak2020deep}
\bibinfo{author}{Wo{\'z}niak, S.}, \bibinfo{author}{Pantazi, A.}, \bibinfo{author}{Bohnstingl, T.} \& \bibinfo{author}{Eleftheriou, E.}
\newblock \bibinfo{journal}{\bibinfo{title}{Deep learning incorporating biologically inspired neural dynamics and in-memory computing}}.
\newblock {\emph{\JournalTitle{Nature Machine Intelligence}}} \textbf{\bibinfo{volume}{2}}, \bibinfo{pages}{325--336} (\bibinfo{year}{2020}).

\bibitem{mehonic2022brain}
\bibinfo{author}{Mehonic, A.} \& \bibinfo{author}{Kenyon, A.~J.}
\newblock \bibinfo{journal}{\bibinfo{title}{Brain-inspired computing needs a master plan}}.
\newblock {\emph{\JournalTitle{Nature}}} \textbf{\bibinfo{volume}{604}}, \bibinfo{pages}{255--260} (\bibinfo{year}{2022}).

\bibitem{paolicelli2011synaptic}
\bibinfo{author}{Paolicelli, R.~C.} \emph{et~al.}
\newblock \bibinfo{journal}{\bibinfo{title}{Synaptic pruning by microglia is necessary for normal brain development}}.
\newblock {\emph{\JournalTitle{science}}} \textbf{\bibinfo{volume}{333}}, \bibinfo{pages}{1456--1458} (\bibinfo{year}{2011}).

\bibitem{faust2021mechanisms}
\bibinfo{author}{Faust, T.~E.}, \bibinfo{author}{Gunner, G.} \& \bibinfo{author}{Schafer, D.~P.}
\newblock \bibinfo{journal}{\bibinfo{title}{Mechanisms governing activity-dependent synaptic pruning in the developing mammalian cns}}.
\newblock {\emph{\JournalTitle{Nature Reviews Neuroscience}}} \textbf{\bibinfo{volume}{22}}, \bibinfo{pages}{657--673} (\bibinfo{year}{2021}).

\bibitem{sellgren2019increased}
\bibinfo{author}{Sellgren, C.~M.} \emph{et~al.}
\newblock \bibinfo{journal}{\bibinfo{title}{Increased synapse elimination by microglia in schizophrenia patient-derived models of synaptic pruning}}.
\newblock {\emph{\JournalTitle{Nature neuroscience}}} \textbf{\bibinfo{volume}{22}}, \bibinfo{pages}{374--385} (\bibinfo{year}{2019}).

\bibitem{knoblauch2010memory}
\bibinfo{author}{Knoblauch, A.}, \bibinfo{author}{Palm, G.} \& \bibinfo{author}{Sommer, F.~T.}
\newblock \bibinfo{journal}{\bibinfo{title}{Memory capacities for synaptic and structural plasticity}}.
\newblock {\emph{\JournalTitle{Neural Computation}}} \textbf{\bibinfo{volume}{22}}, \bibinfo{pages}{289--341} (\bibinfo{year}{2010}).

\bibitem{abbott2004synaptic}
\bibinfo{author}{Abbott, L.} \& \bibinfo{author}{Regehr, W.~G.}
\newblock \bibinfo{journal}{\bibinfo{title}{Synaptic computation}}.
\newblock {\emph{\JournalTitle{Nature}}} \textbf{\bibinfo{volume}{431}}, \bibinfo{pages}{796--803} (\bibinfo{year}{2004}).

\bibitem{wang2020resistive}
\bibinfo{author}{Wang, Z.} \emph{et~al.}
\newblock \bibinfo{journal}{\bibinfo{title}{Resistive switching materials for information processing}}.
\newblock {\emph{\JournalTitle{Nature Reviews Materials}}} \textbf{\bibinfo{volume}{5}}, \bibinfo{pages}{173--195} (\bibinfo{year}{2020}).

\bibitem{huaqiang_2}
\bibinfo{author}{Wei, S.-T.} \emph{et~al.}
\newblock \bibinfo{journal}{\bibinfo{title}{Trends and challenges in the circuit and macro of rram-based computing-in-memory systems}}.
\newblock {\emph{\JournalTitle{Chip}}} \textbf{\bibinfo{volume}{1}}, \bibinfo{pages}{100004} (\bibinfo{year}{2022}).

\bibitem{luwei_2}
\bibinfo{author}{Chen, B.} \emph{et~al.}
\newblock \bibinfo{title}{Efficient in-memory computing architecture based on crossbar arrays}.
\newblock In \emph{\bibinfo{booktitle}{2015 IEEE International Electron Devices Meeting (IEDM)}}, \bibinfo{pages}{17--5} (\bibinfo{organization}{IEEE}, \bibinfo{year}{2015}).

\bibitem{ramanujan2020s}
\bibinfo{author}{Ramanujan, V.}, \bibinfo{author}{Wortsman, M.}, \bibinfo{author}{Kembhavi, A.}, \bibinfo{author}{Farhadi, A.} \& \bibinfo{author}{Rastegari, M.}
\newblock \bibinfo{title}{What's hidden in a randomly weighted neural network?}
\newblock In \emph{\bibinfo{booktitle}{Proceedings of the IEEE/CVF conference on computer vision and pattern recognition}}, \bibinfo{pages}{11893--11902} (\bibinfo{year}{2020}).

\bibitem{LTH_2}
\bibinfo{author}{Pensia, A.}, \bibinfo{author}{Rajput, S.}, \bibinfo{author}{Nagle, A.}, \bibinfo{author}{Vishwakarma, H.} \& \bibinfo{author}{Papailiopoulos, D.}
\newblock \bibinfo{journal}{\bibinfo{title}{Optimal lottery tickets via subset sum: Logarithmic over-parameterization is sufficient}}.
\newblock {\emph{\JournalTitle{Advances in neural information processing systems}}} \textbf{\bibinfo{volume}{33}}, \bibinfo{pages}{2599--2610} (\bibinfo{year}{2020}).

\bibitem{li2023input}
\bibinfo{author}{Li, Y.}, \bibinfo{author}{Moitra, A.}, \bibinfo{author}{Geller, T.} \& \bibinfo{author}{Panda, P.}
\newblock \bibinfo{title}{Input-aware dynamic timestep spiking neural networks for efficient in-memory computing}.
\newblock In \emph{\bibinfo{booktitle}{2023 60th ACM/IEEE Design Automation Conference (DAC)}}, \bibinfo{pages}{1--6} (\bibinfo{organization}{IEEE}, \bibinfo{year}{2023}).

\bibitem{li2024seenn}
\bibinfo{author}{Li, Y.}, \bibinfo{author}{Geller, T.}, \bibinfo{author}{Kim, Y.} \& \bibinfo{author}{Panda, P.}
\newblock \bibinfo{journal}{\bibinfo{title}{Seenn: Towards temporal spiking early exit neural networks}}.
\newblock {\emph{\JournalTitle{Advances in Neural Information Processing Systems}}} \textbf{\bibinfo{volume}{36}} (\bibinfo{year}{2024}).

\bibitem{li2023unleashing}
\bibinfo{author}{Li, C.}, \bibinfo{author}{Jones, E.~G.} \& \bibinfo{author}{Furber, S.}
\newblock \bibinfo{title}{Unleashing the potential of spiking neural networks with dynamic confidence}.
\newblock In \emph{\bibinfo{booktitle}{Proceedings of the IEEE/CVF International Conference on Computer Vision}}, \bibinfo{pages}{13350--13360} (\bibinfo{year}{2023}).

\bibitem{salimans2016improved}
\bibinfo{author}{Salimans, T.} \emph{et~al.}
\newblock \bibinfo{journal}{\bibinfo{title}{Improved techniques for training gans}}.
\newblock {\emph{\JournalTitle{Advances in neural information processing systems}}} \textbf{\bibinfo{volume}{29}} (\bibinfo{year}{2016}).

\bibitem{orchard2015converting}
\bibinfo{author}{Orchard, G.}, \bibinfo{author}{Jayawant, A.}, \bibinfo{author}{Cohen, G.~K.} \& \bibinfo{author}{Thakor, N.}
\newblock \bibinfo{journal}{\bibinfo{title}{Converting static image datasets to spiking neuromorphic datasets using saccades}}.
\newblock {\emph{\JournalTitle{Frontiers in neuroscience}}} \textbf{\bibinfo{volume}{9}}, \bibinfo{pages}{159859} (\bibinfo{year}{2015}).

\bibitem{fang2023spikingjelly}
\bibinfo{author}{Fang, W.} \emph{et~al.}
\newblock \bibinfo{journal}{\bibinfo{title}{Spikingjelly: An open-source machine learning infrastructure platform for spike-based intelligence}}.
\newblock {\emph{\JournalTitle{Science Advances}}} \textbf{\bibinfo{volume}{9}}, \bibinfo{pages}{eadi1480} (\bibinfo{year}{2023}).

\bibitem{van2008visualizing}
\bibinfo{author}{Van~der Maaten, L.} \& \bibinfo{author}{Hinton, G.}
\newblock \bibinfo{journal}{\bibinfo{title}{Visualizing data using t-sne.}}
\newblock {\emph{\JournalTitle{Journal of machine learning research}}} \textbf{\bibinfo{volume}{9}} (\bibinfo{year}{2008}).

\bibitem{wang2019situ}
\bibinfo{author}{Wang, Z.} \emph{et~al.}
\newblock \bibinfo{journal}{\bibinfo{title}{In situ training of feed-forward and recurrent convolutional memristor networks}}.
\newblock {\emph{\JournalTitle{Nature Machine Intelligence}}} \textbf{\bibinfo{volume}{1}}, \bibinfo{pages}{434--442} (\bibinfo{year}{2019}).

\bibitem{qiangfei_1}
\bibinfo{author}{Xia, Q.} \& \bibinfo{author}{Yang, J.~J.}
\newblock \bibinfo{journal}{\bibinfo{title}{Memristive crossbar arrays for brain-inspired computing}}.
\newblock {\emph{\JournalTitle{Nature materials}}} \textbf{\bibinfo{volume}{18}}, \bibinfo{pages}{309--323} (\bibinfo{year}{2019}).

\bibitem{li2018analogue}
\bibinfo{author}{Li, C.} \emph{et~al.}
\newblock \bibinfo{journal}{\bibinfo{title}{Analogue signal and image processing with large memristor crossbars}}.
\newblock {\emph{\JournalTitle{Nature electronics}}} \textbf{\bibinfo{volume}{1}}, \bibinfo{pages}{52--59} (\bibinfo{year}{2018}).

\bibitem{DVS128}
\bibinfo{author}{Amir, A.} \emph{et~al.}
\newblock \bibinfo{title}{A low power, fully event-based gesture recognition system}.
\newblock In \emph{\bibinfo{booktitle}{Proceedings of the IEEE conference on computer vision and pattern recognition}}, \bibinfo{pages}{7243--7252} (\bibinfo{year}{2017}).

\bibitem{kamata2022fully}
\bibinfo{author}{Kamata, H.}, \bibinfo{author}{Mukuta, Y.} \& \bibinfo{author}{Harada, T.}
\newblock \bibinfo{title}{Fully spiking variational autoencoder}.
\newblock In \emph{\bibinfo{booktitle}{Proceedings of the AAAI Conference on Artificial Intelligence}}, vol.~\bibinfo{volume}{36}, \bibinfo{pages}{7059--7067} (\bibinfo{year}{2022}).

\bibitem{yeh2017semantic}
\bibinfo{author}{Yeh, R.~A.} \emph{et~al.}
\newblock \bibinfo{title}{Semantic image inpainting with deep generative models}.
\newblock In \emph{\bibinfo{booktitle}{Proceedings of the IEEE conference on computer vision and pattern recognition}}, \bibinfo{pages}{5485--5493} (\bibinfo{year}{2017}).

\bibitem{peng2021generating}
\bibinfo{author}{Peng, J.}, \bibinfo{author}{Liu, D.}, \bibinfo{author}{Xu, S.} \& \bibinfo{author}{Li, H.}
\newblock \bibinfo{title}{Generating diverse structure for image inpainting with hierarchical vq-vae}.
\newblock In \emph{\bibinfo{booktitle}{Proceedings of the IEEE/CVF Conference on Computer Vision and Pattern Recognition}}, \bibinfo{pages}{10775--10784} (\bibinfo{year}{2021}).

\bibitem{lecun1998gradient}
\bibinfo{author}{LeCun, Y.}, \bibinfo{author}{Bottou, L.}, \bibinfo{author}{Bengio, Y.} \& \bibinfo{author}{Haffner, P.}
\newblock \bibinfo{journal}{\bibinfo{title}{Gradient-based learning applied to document recognition}}.
\newblock {\emph{\JournalTitle{Proceedings of the IEEE}}} \textbf{\bibinfo{volume}{86}}, \bibinfo{pages}{2278--2324} (\bibinfo{year}{1998}).

\bibitem{barratt2018note}
\bibinfo{author}{Barratt, S.} \& \bibinfo{author}{Sharma, R.}
\newblock \bibinfo{journal}{\bibinfo{title}{A note on the inception score}}.
\newblock {\emph{\JournalTitle{arXiv preprint arXiv:1801.01973}}}  (\bibinfo{year}{2018}).

\bibitem{Fashion}
\bibinfo{author}{Xiao, H.}, \bibinfo{author}{Rasul, K.} \& \bibinfo{author}{Vollgraf, R.}
\newblock \bibinfo{journal}{\bibinfo{title}{Fashion-mnist: a novel image dataset for benchmarking machine learning algorithms}}.
\newblock {\emph{\JournalTitle{arXiv preprint arXiv:1708.07747}}}  (\bibinfo{year}{2017}).

\bibitem{lin2020three}
\bibinfo{author}{Lin, P.} \emph{et~al.}
\newblock \bibinfo{journal}{\bibinfo{title}{Three-dimensional memristor circuits as complex neural networks}}.
\newblock {\emph{\JournalTitle{Nature Electronics}}} \textbf{\bibinfo{volume}{3}}, \bibinfo{pages}{225--232} (\bibinfo{year}{2020}).

\bibitem{sebastian2019computational}
\bibinfo{author}{Sebastian, A.}, \bibinfo{author}{Le~Gallo, M.} \& \bibinfo{author}{Eleftheriou, E.}
\newblock \bibinfo{journal}{\bibinfo{title}{Computational phase-change memory: Beyond von neumann computing}}.
\newblock {\emph{\JournalTitle{Journal of Physics D: Applied Physics}}} \textbf{\bibinfo{volume}{52}}, \bibinfo{pages}{443002} (\bibinfo{year}{2019}).

\bibitem{sebastian2020memory}
\bibinfo{author}{Sebastian, A.}, \bibinfo{author}{Le~Gallo, M.}, \bibinfo{author}{Khaddam-Aljameh, R.} \& \bibinfo{author}{Eleftheriou, E.}
\newblock \bibinfo{journal}{\bibinfo{title}{Memory devices and applications for in-memory computing}}.
\newblock {\emph{\JournalTitle{Nature nanotechnology}}} \textbf{\bibinfo{volume}{15}}, \bibinfo{pages}{529--544} (\bibinfo{year}{2020}).

\bibitem{wu2019direct}
\bibinfo{author}{Wu, Y.} \emph{et~al.}
\newblock \bibinfo{title}{Direct training for spiking neural networks: Faster, larger, better}.
\newblock In \emph{\bibinfo{booktitle}{Proceedings of the AAAI conference on artificial intelligence}}, vol.~\bibinfo{volume}{33}, \bibinfo{pages}{1311--1318} (\bibinfo{year}{2019}).

\bibitem{bengio2013estimating}
\bibinfo{author}{Bengio, Y.}, \bibinfo{author}{L{\'e}onard, N.} \& \bibinfo{author}{Courville, A.}
\newblock \bibinfo{journal}{\bibinfo{title}{Estimating or propagating gradients through stochastic neurons for conditional computation}}.
\newblock {\emph{\JournalTitle{arXiv preprint arXiv:1308.3432}}}  (\bibinfo{year}{2013}).

\end{thebibliography}

\end{document}